\documentclass[12pt,draftcls,onecolumn,journal]{IEEEtran}
\usepackage{amsfonts}
\usepackage{amsfonts}
\usepackage{setspace}

\usepackage{amsmath}
\usepackage{amssymb}
\usepackage[dvips]{graphicx}
\usepackage{epsfig}

\newcommand{\snr}{{\footnotesize{SNR}}}
\newcommand{\tsnr}{{\text{\footnotesize{SNR}}}}
\newcommand{\tmin}{\text{min}}
\newcommand{\E}{\mathbb{E}}
\newcommand{\C}{{\sf{C}}}
\newcommand{\Pb}{\bar{P}}
\newcommand{\sta}{{\alpha^{\ast}}}

\newcommand{\tmax}{\text{max}}

\newtheorem{Lem1}{Lemma}
\newtheorem{Lem}{Theorem}

\newtheorem{Exm}{Example}

\begin{document}


\title{Analysis of Energy Efficiency in Fading Channels under QoS Constraints}



%
{\small{\author{\authorblockN{ Mustafa Cenk Gursoy \quad Deli Qiao
\quad Senem Velipasalar}\\ \vspace{0.3cm}
\authorblockA{Department of Electrical Engineering\\
University of Nebraska-Lincoln, Lincoln, NE 68588\\
Email: gursoy@engr.unl.edu, qdl726@bigred.unl.edu,
velipasa@engr.unl.edu}}}}


\maketitle

\thispagestyle{empty}

\begin{abstract} \footnote{The material in this paper was presented
in part at the IEEE Global Communications Conference (Globecom), New
Orleans, in Dec. 2008.
\\
\indent This work was supported by the National Science Foundation under Grants CCF -- 0546384 (CAREER) and CNS -- 0834753. } Energy efficiency in fading channels in the presence
of Quality of Service (QoS) constraints is studied. Effective
capacity, which provides the maximum arrival rate that a wireless
channel can sustain while satisfying statistical QoS constraints, is
considered. Spectral
efficiency--bit energy 
tradeoff is analyzed in the low-power and wideband regimes by
employing the effective capacity formulation, rather than the
Shannon capacity. Through this analysis, energy requirements under
QoS constraints are identified. The analysis is conducted under two
assumptions: perfect channel side information (CSI) available only
at the receiver and perfect CSI available at both the receiver and
transmitter. In particular, it is shown in the low-power regime that
the minimum bit energy required under QoS constraints is the same as
that attained when there are no such limitations. However, this
performance is achieved as the transmitted power vanishes. Through
the wideband slope analysis, the increased energy requirements at
low but nonzero power levels in the presence of QoS constraints are
determined. A similar analysis is also conducted in the wideband
regime, and minimum bit energy and wideband slope expressions are
obtained. In this regime, the required bit energy levels are found
to be strictly greater than those achieved when Shannon capacity is
considered. 
 Overall, a characterization of the energy-bandwidth-delay tradeoff is provided.


\end{abstract}

\newpage

\setcounter{page}{1}

\begin{spacing}{1.8}
\section{Introduction}
Next generation wireless systems will be designed to provide
high-data-rate communications anytime, anywhere in a reliable and
robust fashion while making efficient use of resources. This
wireless vision will enable mobile multimedia communications.
Indeed, one of the features of fourth generation (4G) wireless
systems is the ability to support multimedia services at low
transmission cost \cite[Chap. 23, available online]{Garg}. However,
before this vision is realized, many technical challenges have to be
addressed. In most wireless systems, spectral efficiency and energy
efficiency are important considerations. Especially in mobile
applications, energy resources are scarce and have to be conserved.
Additionally, supporting quality of service (QoS) guarantees is one
of the key requirements in the development of next generation
wireless communication networks. For instance, in real-time services
like multimedia video conference and live broadcast of sporting
events, the key QoS metric is delay. In such cases, information has
to be communicated with minimal delay. Satisfying the QoS
requirements is especially challenging in wireless systems because
channel conditions and hence, for instance, the data rates at which
reliable communication can be established, vary randomly over time
due to mobility and changing environment. Under such volatile
conditions, providing deterministic QoS guarantees either is not
possible or, when it is possible, requires the system to operate
overly pessimistically and achieve low performance underutilizing
the resources. Hence, supporting statistical QoS guarantees is
better suited to wireless systems. In summary, the central issue in
wireless systems is to provide the best performance levels while
satisfying the statistical QoS constraints and making efficient use
of resources.


Information theory provides the ultimate performance limits and
identifies the most efficient use of resources. Due to this fact,
wireless fading channels have been extensively studied from an
information-theoretic point of view, considering different
assumptions on the availability of the channel side information
(CSI) at the receiver and transmitter (see \cite{Biglieri} and
references therein). As also noted above, efficient use of limited
energy resources is of paramount importance in most wireless
systems. From an information-theoretic perspective, the energy
required to reliably send one bit is a metric that can be adopted to
measure the energy efficiency. Generally, energy-per-bit requirement
is minimized, and hence the energy efficiency is maximized, if the
system operates in the low-power or wideband regime. Recently,
Verd\'u in \cite{sergio} has determined the minimum bit energy
required for reliable communications over a general class of
channels, and studied the spectral efficiency--bit energy tradeoff
in the wideband regime. This work has provided a quantitative
analysis of the energy-bandwidth tradeoff.

While providing powerful results, information-theoretic studies
generally do not address delay and QoS constraints
\cite{Ephremides}. For instance, results on the channel capacity
give insights on the performance levels achieved when the
blocklength of codes becomes large \cite{Cover}. The impact upon the
queue length and queueing delay of transmission using codes with
large blocklength can be significant. Situation is even further
exacerbated in wireless channels in which the ergodic capacity has
an operational meaning only if the codewords are long enough to span
all fading states. Now, we also have dependence on fading, and in
slow fading environments, large delays can be experienced in order
to achieve the ergodic capacity. Due to these considerations,
performance metrics such as capacity versus outage \cite{Ozarow} and
delay limited capacity \cite{delay} have been considered in the
literature for slow fading scenarios. For a given outage probability
constraint, outage capacity gives the maximum transmission rate that
satisfies the outage constraint. Delay-limited capacity is defined
as the outage capacity associated with zero outage probability, and
is a performance level that can be attained regardless of the values
of the fading states. Hence, delay limited capacity can be seen as a
deterministic service guarantee. However, delay limited capacity can
be low or even zero, for instance in Rayleigh fading channels even
if both the receiver and transmitter has perfect channel side
information.

More recently, delay constraints are more explicitly considered and
their impact on communication over fading channels is analyzed in
\cite{Berry} and \cite{Neely}. In these studies, the tradeoff
between the  average transmission power and average delay is
identified. More specifically, the authors investigated optimal
transmission and power adaptation policies that take into account
arrival state, buffer occupancy, channel state jointly together and
have the goal of minimizing the average transmission power subject
to average delay constraints.


In this paper, we follow a different approach, and consider
statistical QoS constraints and study the energy efficiency under
such limitations. For this analysis, we employ the notion of
effective capacity \cite{dapeng}, which can be seen as the maximum
throughput that can be achieved by the given energy levels while
providing statistical QoS guarantees. Effective capacity formulation
uses the large deviations theory and incorporates the statistical
QoS constraints by capturing the rate of decay of the buffer
occupancy probability for large queue lengths. In this paper, to
measure the energy efficiency, we consider the bit energy which is
defined as the average energy normalized by the effective capacity.
We investigate the attainable bit energy levels in the low-power and
wideband regimes. For constant source arrival rates, our analysis
provides a tradeoff characterization between the energy and delay.



The rest of the paper is organized as follows. Section II briefly
discusses the system model. Section III reviews the concept of
effective capacity with statistical QoS guarantees, and the spectral
efficiency-bit energy tradeoff. In Section IV, energy efficiency in
the low-power regime is analyzed. Section V investigates the energy
efficiency in the wideband regime.
Finally, Section VI concludes the paper. \vspace{-.3cm}
\section{System Model}

\begin{figure}
\begin{center}
\includegraphics[width=0.65\textwidth]{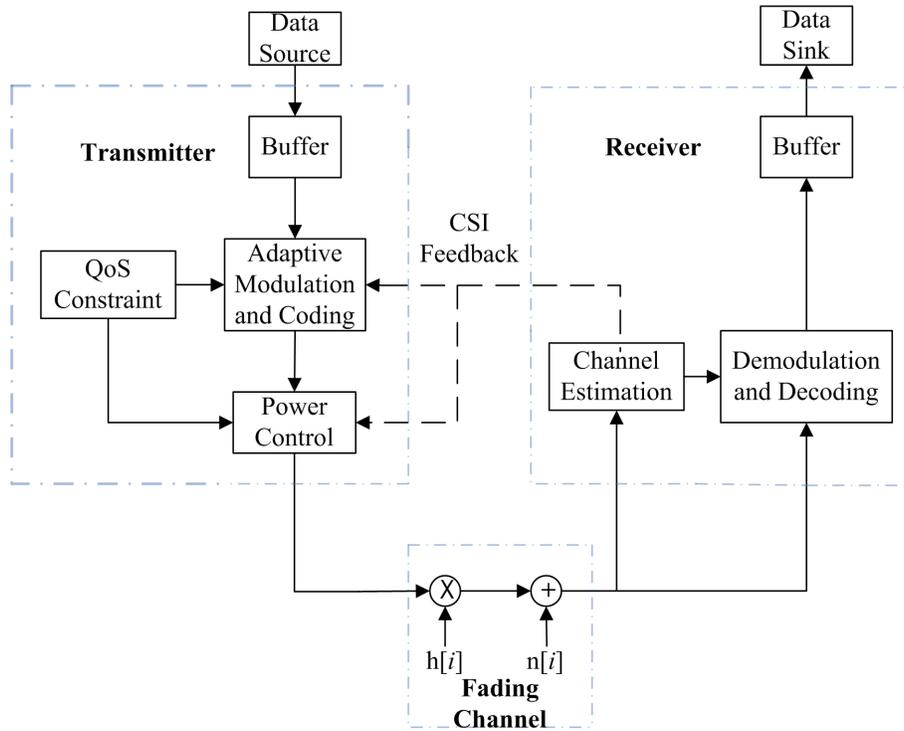}
\caption{The system model}\label{fig:1}
\end{center}
\end{figure}

We consider a point-to-point communication system in which there is
one source and one destination. The general system model is depicted
in Fig.\ref{fig:1}, and is similar to the one studied in \cite{jia}.
In this model, it is assumed that the source generates data
sequences which are divided into frames of duration $T$. These data
frames are initially stored in the buffer before they are
transmitted over the wireless channel. The discrete-time channel
input-output relation in the $i^{\text{th}}$ symbol duration is
given by
\begin{gather} \label{eq:model}
y[i] = h[i] x[i] + n[i] \quad i = 1,2,\ldots.
\end{gather}
where $x[i]$ and $y[i]$ denote the complex-valued channel input and
output, respectively. The channel input is subject to an average
power constraint $\E\{|x[i]|^2\} \le \Pb$ for all $i$, and we assume
that the bandwidth available in the system is $B$. Above, $n[i]$ is
a zero-mean, circularly symmetric, complex Gaussian random variable
with variance $\E\{|n[i]|^2\} = N_0$. The additive Gaussian noise
samples $\{n[i]\}$ are assumed to form an independent and
identically distributed (i.i.d.) sequence. Finally, $h[i]$ denotes
the channel fading coefficient, and $\{h[i]\}$ is a stationary and
ergodic discrete-time process. We assume that perfect channel state
information (CSI) is available at the receiver while the transmitter
has either \emph{no} or \emph{perfect} CSI. The availability of CSI
at the transmitter is facilitated through CSI feedback from the
receiver. Note that if the transmitter knows the channel fading
coefficients, it employs power and rate adaptation. Otherwise, the
signals are sent with constant power.


Note that in the above system model, the average transmitted
signal-to-noise ratio is $\tsnr = \Pb/(N_{0}B)$. We denote the
magnitude-square of the fading coefficient by $z[i]=|h[i]|^2$, and
its distribution function by $p_z(z)$. When there is only receiver
CSI, instantaneous transmitted power is $P[i] = \Pb$ and the
instantaneous received $\tsnr$ is expressed as $\gamma[i]=\Pb
z[i]/(N_{0}B)$. Moreover, the maximum instantaneous service rate
$R[i]$ is \vspace{-.3cm}
\begin{equation}\label{rxrate}
R[i]=B\log_{2}\Big(1+\tsnr z[i]\Big) \quad \text{bits/s}.
\end{equation}
We note that although the transmitter does not know $z[i]$, recently
developed rateless codes such as LT \cite{Luby} and Raptor
\cite{Shok} codes enable the transmitter to adapt its rate to the
channel realization and achieve $R[i]$ without requiring CSI at the
transmitter side \cite{Castura1}, \cite{Castura2}.

When also the transmitter has CSI, the instantaneous service rate is
\begin{equation}\label{txrxrate}
R[i]=B\log_{2}\Big(1+\mu_{\textrm{opt}}(\theta,z[i])z[i]\Big) \quad
\text{bits/s}.
\end{equation}
where $\mu_{\textrm{opt}}(\theta,z)$ is the optimal power-adaptation
policy. The power policy that maximizes the effective capacity,
which will be discussed in Section \ref{subsec:effectivecap}, is
determined in ~\cite{jia}:
\begin{align}\label{power}
\mu_{\textrm{opt}}(\theta,z)=
\begin{cases}
\frac{1}{\alpha^{\frac{1}{\beta+1}}z^{\frac{\beta}{\beta+1}}}-\frac{1}{z}
&z\geq\alpha \\
0 &z<\alpha
\end{cases}
\end{align}
where $\beta=\frac{\theta TB}{\log_e{2}}$ is the normalized QoS
exponent and $\alpha$ is the channel threshold chosen to satisfy the
average power constraint:
\begin{equation}\label{pcon}
\tsnr=\E\{\mu_{\textrm{opt}}(\theta,z)\}=\E\bigg\{\bigg[\frac{1}{\alpha^{\frac{1}{\beta+1}}z^{\frac{\beta}{\beta+1}}}-\frac{1}{z}\bigg]\tau(\alpha)\bigg\}.
\end{equation}
where $\tau(\alpha)=1\{z\geq\alpha\} = \left\{
\begin{array}{ll}
1 & \text{if } z \ge \alpha
\\
0 & \text{if } z < \alpha
\end{array}\right.
$ is the indicator function.

\section{Preliminaries}

In this section, we briefly explain the notion of effective capacity
and also describe the spectral efficiency-bit energy tradeoff. We
refer the reader to \cite{dapeng} and \cite{dapengw} for more
detailed exposition of the effective capacity.

\subsection{Effective Capacity} \label{subsec:effectivecap}

Satisfying quality of service (QoS) requirements is crucial for the
successful deployment and operation of most communication networks.
Hence, in the networking literature, how to handle and satisfy QoS
constraints has been one of the key considerations for many years.
In addressing this issue, the theory of effective bandwidth of a
time-varying source has been developed to identify the minimum
amount of transmission rate that is needed to satisfy the
statistical QoS requirements (see e.g., \cite{Kelly}, \cite{chang},
\cite{changEB}, and \cite{Changbook}).

In wireless communications, the instantaneous channel capacity
varies randomly depending on the channel conditions. Hence, in
addition to the source, the transmission rates for reliable
communication are also time-varying. The time-varying channel
capacity can be incorporated into the theory of effective bandwidth
by regarding the channel service process as a time-varying source
with negative rate and using the source multiplexing rule
(\cite[Example 9.2.2]{Changbook}). Using a similar approach, Wu and
Negi in \cite{dapeng} defined the effective capacity as a dual
concept to effective bandwidth. The effective capacity provides the
maximum constant arrival rate\footnote{Additionally, if the arrival
rates are time-varying, effective capacity specifies the effective
bandwidth of an arrival process that can be supported by the
channel.} that a given time-varying service process can support
while satisfying a QoS requirement specified by $\theta$. If we
define $Q$ as the stationary queue length, then $\theta$ is the
decay rate of the tail
distribution of the queue length: 
\begin{equation}
\lim_{q \to \infty} \frac{\log P(Q \ge q)}{q} = -\theta.
\end{equation}
Therefore, for large $q_{\max}$, we have the following approximation
for the buffer violation probability: $P(Q \ge q_{\max}) \approx
e^{-\theta q_{max}}$. Hence, while larger $\theta$ corresponds to
more strict QoS constraints, smaller $\theta$ implies looser QoS
guarantees. Similarly, if $D$ denotes the steady-state delay
experienced in the buffer, then $P(D \ge d_{\max}) \approx
e^{-\theta \delta d_{\max}}$ for large $d_{\max}$, where $\delta$ is
determined by the arrival and service processes
\cite{tangzhangcross2}. The analysis and application of effect
capacity in various settings has attracted much interest recently
(see e.g., \cite{dapengw}--\cite{finite}).

Let $\{R[i], i=1,2,\ldots\}$ denote the discrete-time stationary and
ergodic stochastic service process and $S[t]\triangleq
\sum_{i=1}^{t}R[i]$ be the time-accumulated process. Assume that the
G\"{a}rtner-Ellis limit of $S[t]$, expressed as \cite{chang}
\begin{equation}
\Lambda_{C}(\theta)=\lim_{t\rightarrow\infty}
\frac{1}{t}\log_e{\mathbb{E}\{e^{\theta S[t]}\}}.
\end{equation}
exists. Then, the effective capacity  is given by ~\cite{dapeng}
\begin{equation}
C_E(\tsnr,\theta)=-\frac{\Lambda_{C}(-\theta)}{\theta}=-\lim_{t\rightarrow\infty}\frac{1}{\theta
t}\log_e{\mathbb{E}\{e^{-\theta S[t]}\}}.
\end{equation}
If the fading process $\{h[i]\}$ is constant during the frame
duration $T$ and changes independently from frame to frame,  then
the effective capacity simplifies to
\begin{equation}\label{ec}
C_{E}(\tsnr,\theta)=-\frac{1}{\theta T}\log_e\mathbb{E}\{e^{-\theta
T R[i]}\} \quad \text{bits/s}.
\end{equation}
This block-fading assumption is an approximation for practical
wireless channels, and the independence assumption can be justified
if, for instance, transmitted frames are interleaved before
transmission, or time-division multiple access is employed and frame
duration is proportional to the coherence time of the channel.

It can be easily shown that effective capacity specializes to the
Shannon capacity and delay-limited capacity in the asymptotic
regimes. As $\theta$ approaches to 0, constraints on queue length
and queueing delay relax, and effective capacity converges to the
Shannon ergodic capacity:
\begin{gather}
\lim_{\theta \to 0} C_{E}(\tsnr,\theta) = \left\{
\begin{array}{ll}
\E\{B\log_{2}(1+\tsnr z)\} & \text{CSI at the RX}
\\
\E\left\{B\log_{2}\left(1+\mu_{\textrm{opt}}(\theta,z)z\right)\right\}
& \text{CSI at the RX and TX}
\end{array}\right.
\end{gather}
where expectations are with respect to $z$. On the other hand, as
$\theta \to \infty$, QoS constraints become more and more strict and
effective capacity approaches the delay-limited capacity which as
described before can be seen as a deterministic service guarantee:
\begin{gather}
\lim_{\theta \to \infty} C_{E}(\tsnr,\theta) = \left\{
\begin{array}{ll}
B\log_{2}(1+\tsnr z_{\min}) & \text{CSI at the RX}
\\
B\log_{2}\left(1 + \sigma \right) & \text{CSI at the RX and TX}
\end{array}\right.
\end{gather}
where $\sigma=\frac{\tsnr}{\E\{1/z\}}$ and $z_{\min}$ is the minimum
value of the random variable $z$, i.e., $z \ge z_{\min} \ge 0$ with
probability 1. Note that in Rayleigh fading, $\sigma = 0$ and
$z_{\min} = 0$, and hence the delay-limited capacities are zero in
both cases and no deterministic guarantees can be provided.

\subsection{Spectral Efficiency vs. Bit Energy} \label{subsec:seeb}

In \cite{sergio}, Verd\'u has extensively studied the spectral
efficiency--bit energy tradeoff in the wideband regime. In this
work, the minimum bit energy required for reliable communication
over a general class of multiple-input multiple-output channels is
identified. In general, if the capacity is a concave function of
$\tsnr$, then the minimum bit energy is achieved as $\tsnr \to 0$.
Additionally, Verd\'u has defined the wideband slope, which is the
slope of the spectral efficiency curve at zero spectral efficiency.
While the minimum bit energy is a performance measure as $\tsnr \to
0$, wideband slope has emerged as a tool that enables us to analyze
the energy efficiency at low but nonzero power levels and at large
but finite bandwidths. In \cite{sergio}, the tradeoff between
spectral efficiency and energy efficiency is analyzed considering
the Shannon capacity. In this paper, we perform a similar analysis
employing the effective capacity. Here, we denote the effective
capacity normalized by bandwidth or equivalently the spectral
efficiency in bits per second per Hertz by
\begin{gather} \label{eq:specteff}
\C_E(\tsnr,\theta)=\frac{C_{E}(\tsnr,\theta)}{B} = -\frac{1}{\theta
T B}\log_e\mathbb{E}\{e^{-\theta T R[i]}\}.
\end{gather}
Hence, we characterize the spectral efficiency--bit energy tradeoff
under QoS constraints. Note that effective capacity provides a
characterization of the arrival process. However, since the average
arrival rate is equal to the average departure rate when the queue
is in steady-state \cite{ChangZajic}, effective capacity can also be
seen as a measure of the average rate of transmission. We first have
the following preliminary result.

\begin{Lem1}
The normalized effective capacity, $\C_{E}(\tsnr)$, given in
(\ref{eq:specteff}) is a concave function of $\tsnr$.
\end{Lem1}

\emph{Proof}: It can be easily seen that $e^{-\theta T R[i]}$, where
$R[i] = B\log_{2}(1+\tsnr z[i])$ or
$R[i]=B\log_{2}(1+\mu_{\textrm{opt}}(\theta,z[i])z[i])$, is a
log-convex function of $\tsnr$ because $-R[i]$ is a convex function
of $\tsnr$. Since log-convexity is preserved under sums, $g(x) =
\int f(x,y) dy$ is log-convex in $x$ if $f(x,y)$ is log-convex in
$x$ for each $y$ \cite{convex}. From this fact, we immediately
conclude that $\E\{e^{-\theta T R[i]}\}$ is also a log-convex
function of $\tsnr$. Hence, $\log_e \E\{e^{-\theta T R[i]}\}$ is
convex and $-\log_e \E\{e^{-\theta T R[i]}\}$ is concave in
$\tsnr$.\hfill $\square$

Then, it can be easily seen that $\frac{E_b}{N_0}_{\tmin}$ under QoS
constraints can be obtained from \cite{sergio}
\begin{equation}\label{ebmin}
\frac{E_b}{N_0}_{\tmin}=\lim_{\tsnr\rightarrow0}\frac{\tsnr}{\C_{E}(\tsnr)}=\frac{1}{\dot{\C}_{E}(0)}.
\end{equation}
At $\frac{E_b}{N_0}_{\tmin}$, the slope $\mathcal {S}_0$ of the
spectral efficiency versus $E_b/N_0$ (in dB) curve is defined as
\cite{sergio}
\begin{equation}\label{slope}
\mathcal{S}_0=\lim_{\frac{E_b}{N_0}\downarrow\frac{E_b}{N_0}_\tmin}
\frac{\C_E(\frac{E_b}{N_0})}{10\log_{10}\frac{E_b}{N_0}-10\log_{10}\frac{E_b}{N_0}_\tmin}10\log_{10}2.
\end{equation}
Considering the expression for normalized effective capacity, the
wideband slope can be found from
\begin{equation}\label{ecslope}
\mathcal{S}_0=-\frac{2(\dot{\C}_E(0))^2}{\ddot{\C}_E(0)}\log_e{2}
\end{equation}
where $\dot{\C}_E(0)$ and $\ddot{\C}_E(0)$ are the first and second
derivatives, respectively, of the function $\C_E(\tsnr)$ in
bits/s/Hz at zero $\tsnr$ \cite{sergio}. $\frac{E_b}{N_0}_{\tmin}$
and $\mathcal{S}_0$ provide a linear approximation of the spectral
efficiency curve at low spectral efficiencies, i.e.,
\begin{gather}\label{eq:linearapprox}
\C_E\left(\frac{E_b}{N_0}\right) = \frac{\mathcal{S}_0}{10\log_{10}
2} \left(
\frac{E_b}{N_0}\bigg|_{dB}-\frac{E_b}{N_0}_\tmin\bigg|_{dB}\right) +
\epsilon
\end{gather}
where $\frac{E_b}{N_0}\Big|_{dB} = 10\log_{10}\frac{E_b}{N_0}$ and
$\epsilon = o\left( \frac{E_b}{N_0}-\frac{E_b}{N_0}_\tmin\right)$.

\section{Energy Efficiency in the Low-Power Regime}\label{sec:lowp}

As discussed in the previous section, the minimum bit energy is
achieved as $\tsnr = \frac{\Pb}{N_0 B} \to 0$, and hence energy
efficiency improves if one operates in the low-power or
high-bandwidth regime. From the Shannon capacity perspective,
similar performances are achieved in these two regimes, which
therefore can be seen as equivalent. However, as we shall see in
this paper, considering the effective capacity leads to different
results at low power and high bandwidth levels.  In this section, we
consider the low-power regime for fixed bandwidth, $B$, and study
the spectral efficiency vs. bit energy tradeoff by finding the
minimum bit energy and the wideband slope.

\subsection{CSI  at the Receiver Only}
We initially consider the case in which only the receiver knows the
channel conditions. Substituting (\ref{rxrate}) into (\ref{ec}), we
obtain the spectral efficiency given $\theta$ as a function of \snr:
\begin{equation}\label{ecrxsnr}
\C_E(\tsnr)=-\frac{1}{\theta TB}\log_e{\mathbb{E}\{e^{-\theta
TB\log_{2}(1+\tsnr z)}\}}=-\frac{1}{\theta
TB}\log_e{\mathbb{E}\{(1+\tsnr z)^{-\beta}\}}
\end{equation}
where again $\beta = \frac{\theta T B}{\log_e 2}$. Note that since
the analysis is performed for fixed $\theta$ throughout the paper,
we henceforth express the effective capacity only as a function of
$\tsnr$ to simplify the expressions. The following result provides
the minimum bit energy and the wideband slope.
\begin{Lem} \label{theo:lowpowercsir}
When only the receiver has perfect CSI, the minimum bit energy and
wideband slope are
\begin{gather}
\frac{E_b}{N_0}_{\tmin}=\frac{\log_e{2}}{\E\{z\}} \text{ and }
\mathcal{S}_0=\frac{2}{(\beta+1)\frac{\E\{z^{2}\}}{\big(\E\{z\}\big)^{2}}-\beta}.
\end{gather}
\end{Lem}
\emph{Proof}: The first and second derivative of $\C_E(\tsnr)$ with
respect to $\tsnr$ are given by
\begin{gather}\label{dotecsnr}
\dot{\C}_E(\tsnr)=\frac{1}{\log_e{2}}\frac{\E\{(1+\tsnr
z)^{-(\beta+1)}z\}}{\E\{(1+\tsnr z)^{-\beta}\}} \quad \text{and,}\\
\ddot{\C}_E(\tsnr)=\frac{\beta}{\log_e{2}}\bigg(\frac{\E\{(1+\tsnr
z)^{-(\beta+1)}z\}}{\E\{(1+\tsnr
z)^{-\beta}\}}\bigg)^2-\frac{\beta+1}{\log_e{2}}\frac{\E\{(1+\tsnr
z)^{-(\beta+2)}z^{2}\}}{\E\{(1+\tsnr z)^{-\beta}\}},
\label{ddotecsnr}
\end{gather}
respectively, which result in the following expressions when $\tsnr
= 0$:
\begin{gather}\label{eq:derivatives}
\dot{\C}_E(0)=\frac{\E\{z\}}{\log_e{2}} \quad \text{and} \quad
\ddot{\C}_E(0)=-\frac{1}{\log_e{2}}\Big((\beta+1)\E\{z^{2}\}-\beta\big(\E\{z\}\big)^{2}\Big).
\end{gather}
Substituting the expressions in (\ref{eq:derivatives}) into
(\ref{ebmin}) and (\ref{ecslope}) provides the desired result.
\hfill $\square$

From the above result, we immediately see that
$\frac{E_b}{N_0}_{\min}$ does not depend on $\theta$ and the minimum
\emph{received} bit energy is
\begin{equation}\label{ebr}
\frac{E_b^r}{N_0}_{\tmin}=\frac{E_b}{N_0}_{\tmin}\E\{z\}=\log_e2=-1.59
\text{ dB}.
\end{equation}
Note that if the Shannon capacity is used in the analysis, i.e., if
$\theta = 0$ and hence $\beta = 0$, $\frac{E_b^r}{N_0}_{\tmin} =
-1.59$ dB and $\mathcal{S}_0=2/(\E\{z^{2}\}/\E^2\{z\})$. Therefore,
we conclude from Theorem \ref{theo:lowpowercsir} that as the average
power $\Pb$ decreases, energy efficiency approaches the performance
achieved by a system that does not have QoS limitations. However, we
note that wideband slope is smaller if $\theta > 0$. Hence, the
presence of QoS constraints decreases the spectral efficiency or
equivalently increases the energy requirements for fixed spectral
efficiency values at low but nonzero $\tsnr$ levels.

\begin{figure}
\begin{center}
\includegraphics[width=0.65\textwidth]{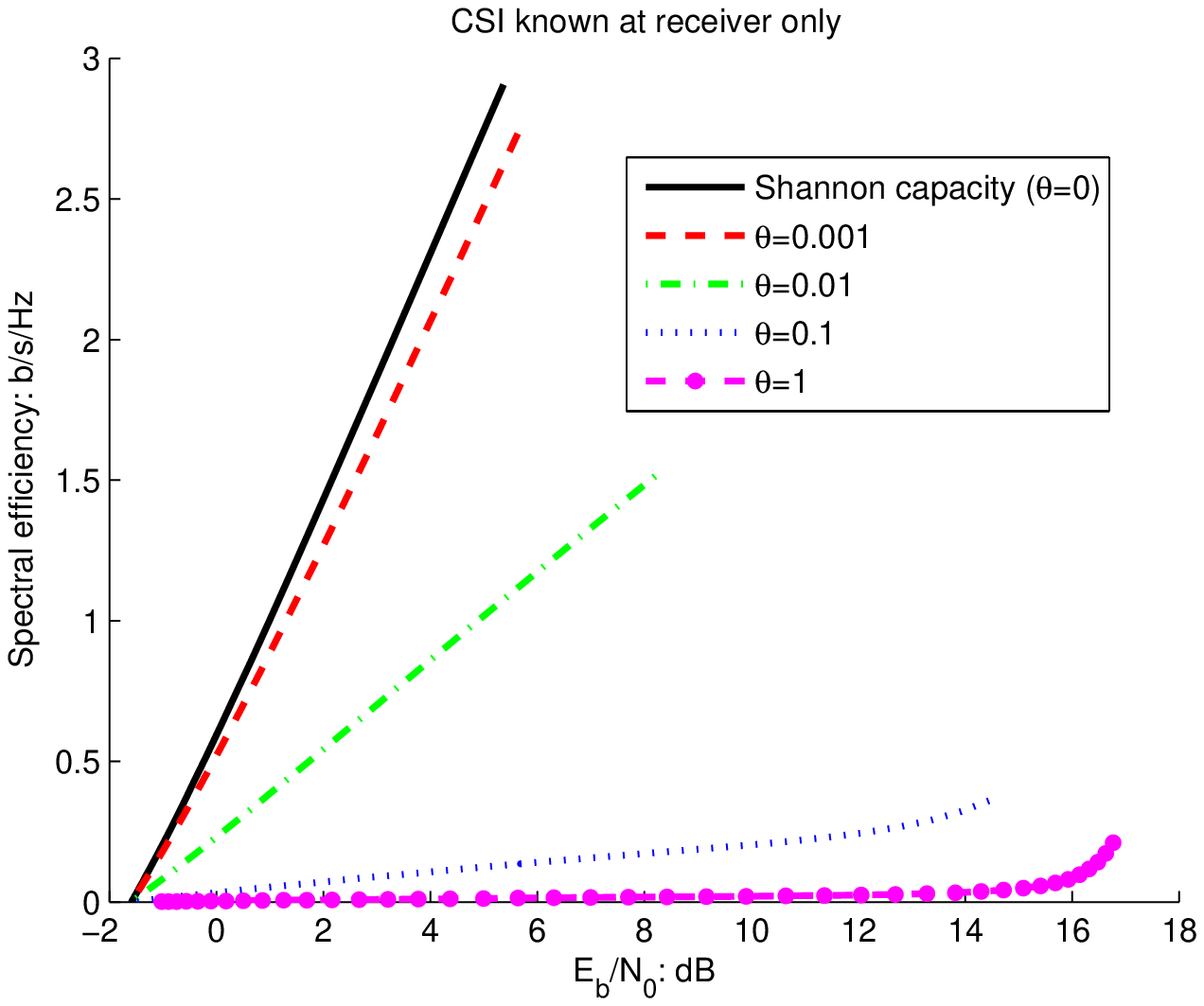}
\caption{Spectral efficiency vs. $E_{b}/N_{0}$ in the Rayleigh
fading channel with fixed $B$; CSI known at the receiver
only.}\label{fig:2}
\end{center}
\end{figure}

Fig. \ref{fig:2} plots the spectral efficiency as a function of the
bit energy for different values of $\theta$ in the Rayleigh fading
channel with $\E\{|h|^2\} = \E\{z\} = 1$.  Note that the curve for
$\theta = 0$ corresponds to the Shannon capacity. Throughout the
paper, we set the frame duration to $T=2$ms in the numerical
results. For the fixed bandwidth case, we have assumed $B=10^5$ Hz.
In Fig. \ref{fig:2}, we observe that all curves approach
$\frac{E_b}{N_0}_\tmin = -1.59$ dB as predicted. On the other hand,
we note that the wideband slope decreases as $\theta$ increases.
Therefore, at low but nonzero spectral efficiencies, more energy is
required as the QoS constraints become more stringent. Considering
the linear approximation in (\ref{eq:linearapprox}), we can easily
show for fixed spectral efficiency $\C\left(\frac{E_b}{N_0}\right)$
for which the linear approximation is accurate that the increase in
the bit energy in dB, when the QoS exponent increases from
$\theta_1$ to $\theta_2$, is
\begin{gather}
\frac{E_b}{N_0}\bigg|_{dB,
\theta_2}-\frac{E_b}{N_0}\bigg|_{dB,\theta_1} =
\left(\frac{1}{\mathcal{S}_{0,\theta_2}} -
\frac{1}{\mathcal{S}_{0,\theta_1}}\right)
\C\left(\frac{E_b}{N_0}\right) 10\log_{10}2.
\end{gather}


\subsection{CSI at both the Transmitter and Receiver}

We now consider the case in which both the transmitter and receiver
have perfect CSI. Substituting (\ref{txrxrate}) into (\ref{ec}), we
have
\begin{equation}\label{ectxrxsnr}
\C_E(\tsnr)=-\frac{1}{\theta TB}\log_e\E\Big\{e^{-\theta
TB\log_{2}\big(1+\mu_{\textrm{opt}}(\theta,z)z\big)}\Big\}=-\frac{1}{\theta
TB}\log_e\Big(F(\alpha)+\E\Big\{\left(\frac{z}{\alpha}\right)^{-\frac{\beta}{\beta+1}}\tau(\alpha)\Big\}\Big)
\end{equation}
where $F(\alpha)=\E\{1\{z<\alpha\}\}$. For this case, following an
approach similar to that in \cite{cdma}, we obtain the following
result.
\begin{Lem} \label{theo:lowpowercsirt}
When both the transmitter and receiver have perfect CSI, the minimum
bit energy with optimal power control and rate adaptation becomes
\begin{gather}
\frac{E_b}{N_0}_{\tmin}= \frac{\log_e 2}{z_{\max}}
\end{gather}
where $z_{\max}$ is the essential supremum of the random variable
$z$, i.e., $z \le z_{\max}$ with probability 1.
\end{Lem}
\emph{Proof}: We assume that $z_{\tmax}$ is the maximum value that
the random variable $z$ can take, i.e., $P(z \le z_\tmax)=1$. From
(\ref{pcon}), we can see that as $\tsnr$ vanishes, $\alpha$
increases to $z_{\max}$, because otherwise while $\tsnr$ approaches
zero, the right most side of (\ref{pcon}) does not. Then, we can
suppose for small enough $\tsnr$ that
\begin{equation}\label{supalpha}
\alpha=z_{\tmax}-\eta
\end{equation}
\end{spacing}
\begin{spacing}{1.7}
where $\eta\rightarrow0$ as $\tsnr\rightarrow0$. Substituting
(\ref{supalpha}) into (\ref{pcon}) and (\ref{ectxrxsnr}), we get
\begin{align}
\frac{E_b}{N_0}_{\tmin}=\lim_{\tsnr\rightarrow0}\frac{\tsnr}{\C(\tsnr)}&=\lim_{\eta\rightarrow0}\frac{\E\bigg\{\Big[\frac{1}{(z_{\tmax}-\eta)^{\frac{1}{\beta+1}}z^{\frac{\beta}{\beta+1}}}-\frac{1}{z}\Big]\tau(z_{\tmax}-\eta)\bigg\}}{-\frac{1}{\theta
TB}\log_e\Big(F(z_{\tmax}-\eta)+\E\Big\{\big(\frac{z}{z_{\tmax}-\eta}\big)^{-\frac{\beta}{\beta+1}}\tau(z_{\tmax}-\eta)\Big\}\Big)} \label{eq:csirtproof0}\\
&=\lim_{\eta\rightarrow0}\frac{\int_{z_{\tmax}-\eta}^{z_{\tmax}}\bigg(\frac{1}{(z_{\tmax}-\eta)^{\frac{1}{\beta+1}}z^{\frac{\beta}{\beta+1}}}-\frac{1}{z}\bigg)p_z(z)dz}{-\frac{1}{\theta
TB}\log_e\bigg(\int_{0}^{z_{\tmax}-\eta}p_z(z)dz+\int_{z_{\tmax}-\eta}^{z_{\tmax}}\big(\frac{z}{z_{\tmax}-\eta}\big)^{-\frac{\beta}{\beta+1}}p_z(z)dz\bigg)} \label{eq:csirtproof1}\\
&=\lim_{\eta\rightarrow0}\frac{\frac{1}{\beta+1}(z_{\tmax}-\eta)^{-\frac{\beta+2}{\beta+1}}\int_{z_{\tmax}-\eta}^{z_{\tmax}}
\frac{p_z(z)}{z^{\frac{\beta}{\beta+1}}}dz}{-\frac{1}{\beta\log_e2}\frac{-\frac{\beta}{\beta+1}(z_{\tmax}-\eta)^{-\frac{1}{\beta+1}}
\int_{z_{\tmax}-\eta}^{z_{\tmax}}\frac{p_z(z)}{z^{\frac{\beta}{\beta+1}}}dz}
{\int_{0}^{z_{\tmax}-\eta}p_z(z)dz+\int_{z_{\tmax}-\eta}^{z_{\tmax}}\big(\frac{z}{z_{\tmax}-\eta}\big)^{-\frac{\beta}{\beta+1}}p_z(z)dz}}\label{eq:csirtproof2}\\
&=\lim_{\eta\rightarrow0}\frac{\Big(\int_{0}^{z_{\tmax}-\eta}p_z(z)dz+\int_{z_{\tmax}-\eta}^{z_{\tmax}}\big(\frac{z}{z_{\tmax}-\eta}\big)^{-\frac{\beta}{\beta+1}}p_z(z)dz\Big)\log_e2}{z_{\tmax}-\eta}=\frac{\log_e2}{z_{\tmax}}\label{eq:csirtproof3}
\end{align}
where $p_z$ is the distribution of channel gain $z$.
(\ref{eq:csirtproof1}) is obtained by expressing the expectations in
(\ref{eq:csirtproof0}) as integrals. (\ref{eq:csirtproof2}) follows
by using the L'Hospital's Rule and applying Leibniz Integral Rule.
The first term in (\ref{eq:csirtproof3}) is obtained after
straightforward algebraic simplifications and the result follows
immediately. \hfill $\square$

Note that for distributions with unbounded support, we have
$z_{\max} = \infty$ and hence $\frac{E_b}{N_0}_{\tmin}= 0 = -\infty$
dB. In this case, it is easy to see that the wideband slope is $S_0
= 0$.
\begin{Exm}\label{example}
Specifically, for the Rayleigh fading channel, as in ~\cite{sha}, it
can be shown that
\\
$\lim_{\tsnr\rightarrow0}{\frac{\C_E(\tsnr)}{\tsnr\log_e(\frac{1}{\tsnr})\log_e{2}}}=1.
$
Then, spectral efficiency can be written as
$\C_E(\tsnr)\approx\tsnr\log_e(\frac{1}{\tsnr})\log_e{2}$, so
$
\frac{E_b}{N_0}_\textrm{min}=\lim_{\tsnr\rightarrow0}\frac{\tsnr}{\C_E(\tsnr)}=\lim_{\tsnr\rightarrow0}\frac{1}{\log_e(\frac{1}{\tsnr})\log_e{2}}=0
$
which also verifies the above result.
\end{Exm}

\emph{Remark:} We note that as in the case in which there is CSI at
the receiver, the minimum bit energy achieved under QoS constraints
is the same as that achieved by the Shannon capacity \cite{cdma}.
Hence, the energy efficiency again approaches the performance of an
unconstrained system as power diminishes. Searching for an intuitive
explanation of this observation, we note that arrival rates that can
be supported vanishes with decreasing power levels. As a result, the
impact of buffer occupancy constraints on the performance lessens.
Note that in contrast, increasing the bandwidth increases the
arrival rates supported by the system. Therefore, limitations on the
buffer occupancy will have significant impact upon the energy
efficiency in the wideband regime as will be discussed in Section
\ref{sec:wideb}.

\begin{figure}
\begin{center}
\includegraphics[width=0.65\textwidth]{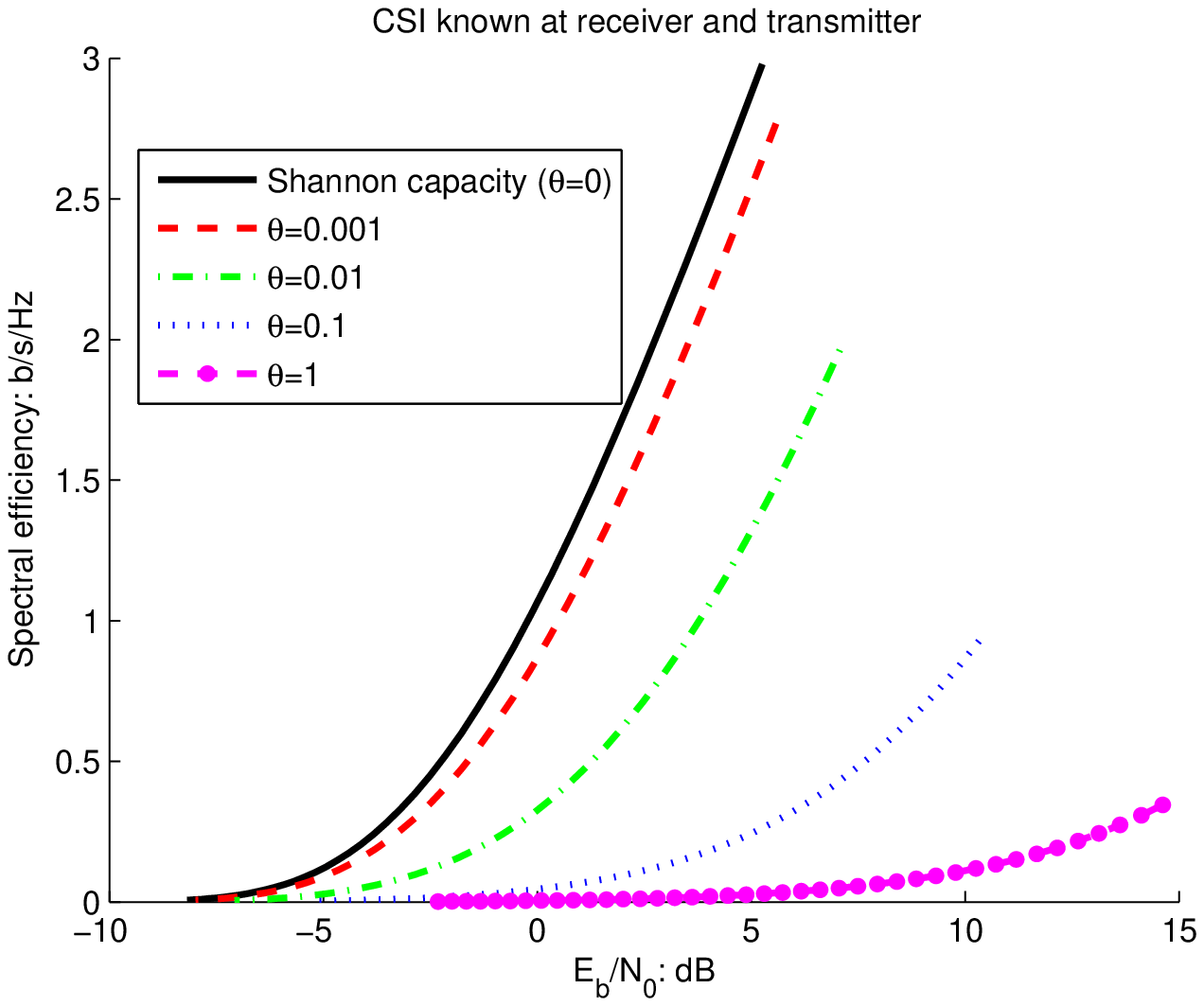}
\caption{Spectral efficiency vs. $E_{b}/N_{0}$ in the Rayleigh
fading channel with fixed $B$; CSI known at the transmitter and
receiver.}\label{fig:4}
\end{center}
\end{figure}

Fig. \ref{fig:4} plots the spectral efficiency vs. bit energy for
different values of $\theta$ in the Rayleigh fading channel with
$\E\{z\} = 1$. In all cases, we observe that the bit energy goes to
$-\infty$ as the spectral efficiency decreases. We also note that at
small but nonzero spectral efficiencies, the required energy is
higher as $\theta$ increases.

\section{Energy Efficiency in the Wideband Regime}\label{sec:wideb}

In this section, we study the performance at high bandwidths while
the average power $\Pb$ is kept fixed. We investigate the impact of
$\theta$ on $\frac{E_b}{N_0}_\tmin$ and the wideband slope
$\mathcal{S}_0$ in this wideband regime. Note that as the bandwidth
increases, the average signal-to-noise ratio $\tsnr=\Pb/(N_{0}B)$
and the spectral efficiency decreases.

\subsection{CSI at the Receiver Only}
We define $\zeta = \frac{1}{B}$ and express the spectral efficiency
(\ref{ecrxsnr}) as a function of $\zeta$:
\begin{equation}\label{rxecb}
\C_E(\zeta)=-\frac{\zeta}{\theta
T}\log_e{\mathbb{E}\{e^{-\frac{\theta
T}{\zeta}\log_{2}(1+\frac{\Pb\zeta}{N_0}z)}\}}.
\end{equation}
The bit energy is again defined as
\begin{gather}
\frac{E_b}{N_0} = \frac{\tsnr}{\C_E(\tsnr)} =
\frac{\frac{\Pb\zeta}{N_0}}{\C_E(\zeta)} =
\frac{\frac{\Pb}{N_0}}{\C_E(\zeta)/\zeta}.
\end{gather}
It can be readily verified that $\C_E(\zeta)/\zeta$ monotonically
increases as $\zeta \to 0$ (or equivalently as $B \to \infty$) (see
Appendix \ref{app:2}). Therefore
\begin{equation}\label{ebminb}
\frac{E_b}{N_0}_{\tmin}=\lim_{\zeta
\rightarrow0}\frac{\Pb\zeta/N_{0}}{\C_{E}(\zeta)}=\frac{\Pb/N_{0}}{\dot{\C}_{E}(0)}
\end{equation}
where $\dot{\C}_E(0)$ is the first derivative of the spectral
efficiency with respect to $\zeta$ at $\zeta = 0$. The wideband
slope $\mathcal{S}_0$ can be obtained from the formula
(\ref{ecslope}) by using the first and second derivatives of the
spectral efficiency $\C_E(\zeta)$ with respect to $\zeta$.
\begin{Lem}
When only the receiver has CSI, the minimum bit energy and wideband
slope, respectively, in the wideband regime are given by
\begin{gather}
\frac{E_b}{N_0}_{\tmin}=\frac{-\frac{\theta
T\Pb}{N_0}}{\log_e\E\{e^{-\frac{\theta T\Pb}{N_{0}\log_2}z}\}},
\quad \text{and} \label{eq:ebminwbr}
\\
\mathcal{S}_{0}=2\Big(\frac{N_{0}\log_e2}{\theta
T\Pb}\Big)^{2}\frac{\E\{e^{-\frac{\theta
T\Pb}{N_{0}\log_e2}z}\}\Big(\log_e\E\{e^{-\frac{\theta
T\Pb}{N_{0}\log_e2}z}\}\Big)^{2}}{\E\{e^{-\frac{\theta
T\Pb}{N_{0}\log_e2}z}z^{2}\}}. \label{eq:slopewbr}
\end{gather}
\end{Lem}

\emph{Proof}: The first and second derivative of $\C_E(\zeta)$ are
given by
\begin{equation}\label{dotecb}
\dot{\C}_E(\zeta)=-\frac{1}{\theta T} \log_e\E\{e^{-\frac{\theta
T}{\zeta}\log_{2}(1+\frac{\Pb\zeta
z}{N_0})}\}-\frac{\E\bigg\{e^{-\frac{\theta T}
{\zeta}\log_{2}(1+\frac{\Pb\zeta
z}{N_0})}\Bigl[\frac{\log_{2}(1+\frac{\Pb\zeta
z}{N_0})}{\zeta}-\frac{\frac{\Pb
z}{N_{0}\log_e{2}}}{1+\frac{\Pb\zeta z}{N_0}}\Bigr]\bigg\}}{\E\{
e^{-\frac{\theta T}{\zeta}\log_{2}(1+\frac{\Pb\zeta z}{N_0})}\}},
\end{equation}
\begin{equation}\label{ddotecb}
\begin{split}
\ddot{\C}_E(\zeta)&=\frac{\theta
T}{\zeta}\left(\frac{\E\bigg\{e^{-\frac{\theta T}
{\zeta}\log_{2}(1+\frac{\Pb\zeta
z}{N_0})}\Bigl[\frac{\log_{2}(1+\frac{\Pb\zeta
z}{N_0})}{\zeta}-\frac{\frac{\Pb
z}{N_{0}\log_e{2}}}{1+\frac{\Pb\zeta z}{N_0}}\Bigr]\bigg\}}{\E\{
e^{-\frac{\theta T}{\zeta}\log_{2}(1+\frac{\Pb\zeta
z}{N_0})}\}}\right)^{2}\\
&\phantom{{=}\frac{\theta T}{\zeta}}-\frac{\E\bigg\{e^{-\frac{\theta
T}{\zeta} \log_{2}(1+\frac{\Pb\zeta z}{N_0})}\biggl[\frac{\theta
T}{\zeta}\Bigl(\frac{\log_{2}(1+\frac{\Pb\zeta
z}{N_0})}{\zeta}-\frac{\frac{\Pb
z}{N_{0}\log_e{2}}}{1+\frac{\Pb\zeta
z}{N_0}}\Bigr)^{2}+\log_e{2}\Big(\frac{\frac{\Pb
z}{N_{0}\log_e{2}}}{1+\frac{\Pb\zeta z}{N_0}}
\Big)^{2}\biggr]\bigg\}}{\E\{ e^{-\frac{\theta
T}{\zeta}\log_{2}(1+\frac{\Pb\zeta z}{N_0})}\}}
\end{split}.
\end{equation}
First, we define the function
$f(\zeta)=\frac{\log_{2}(1+\frac{\Pb\zeta
z}{N_0})}{\zeta^{2}}-\frac{\frac{\Pb
z}{N_{0}\zeta\log_e{2}}}{1+\frac{\Pb\zeta z}{N_0}}. $
Then, we can show that
\begin{equation}\label{temp0}
\begin{split}
\lim_{\zeta\rightarrow0}{f(\zeta)}
=\lim_{\zeta\rightarrow0}{\frac{\frac{\log_{2}(1+\frac{\Pb\zeta
z}{N_0})}{\zeta}-\frac{\frac{\Pb
z}{N_{0}\log_e{2}}}{1+\frac{\Pb\zeta z}{N_0}}}{\zeta}}
&=\lim_{\zeta\rightarrow0}\biggl(-\frac{\log_{2}(1+\frac{\Pb\zeta
z}{N_0})}{\zeta^{2}}+\frac{\frac{\Pb
z}{N_{0}\log_e{2}}}{1+\frac{\Pb\zeta z}{N_0}}+\Bigl(\frac{\frac{\Pb
z}{N_{0}\log_e{2}}}{1+\frac{\Pb\zeta
z}{N_0}}\Bigr)^{2}\log_e{2}\biggr)\nonumber\\
&=-\lim_{\zeta\rightarrow0}{f(\zeta)}+\frac{1}{\log_e{2}}\left(\frac{\Pb z}{N_0}\right)^2\\
\end{split}
\end{equation}
which yields
\begin{equation}\label{temp00}
\lim_{\zeta\rightarrow0}
{f(\zeta)}=\frac{1}{2\log_e{2}}\left(\frac{\Pb
z}{N_0}\right)^2
\end{equation}
Using (\ref{temp00}), we can easily find from (\ref{dotecb}) that
\begin{equation}\label{dotecb0}
\lim_{\zeta\rightarrow0} {\dot{\C}_E(\zeta)}=-\frac{1}{\theta
T}\log_e\E\left\{e^{-\frac{\theta T \Pb}{N_0\log_e{2}}z}\right\}
\end{equation}
from which (\ref{eq:ebminwbr}) follows immediately. Moreover, from
(\ref{ddotecb}), we can derive
\begin{equation}\label{ddotecb0}
\lim_{\zeta\rightarrow0}
{\ddot{\C}_E(\zeta)}=-\frac{1}{\log_e2}\Big(\frac{\Pb}{N_0}\Big)^{2}\frac{\E\{e^{-\frac{\theta
T\Pb}{N_{0}\log_e2}z}z^{2}\}}{\E\{e^{-\frac{\theta
T\Pb}{N_{0}\log_e2}z}\}}.
\end{equation}
Evaluating (\ref{ecslope}) with (\ref{dotecb0}) and (\ref{ddotecb0})
provides (\ref{eq:slopewbr}).\hfill$\square$

\begin{figure}
\begin{center}
\includegraphics[width=0.65\textwidth]{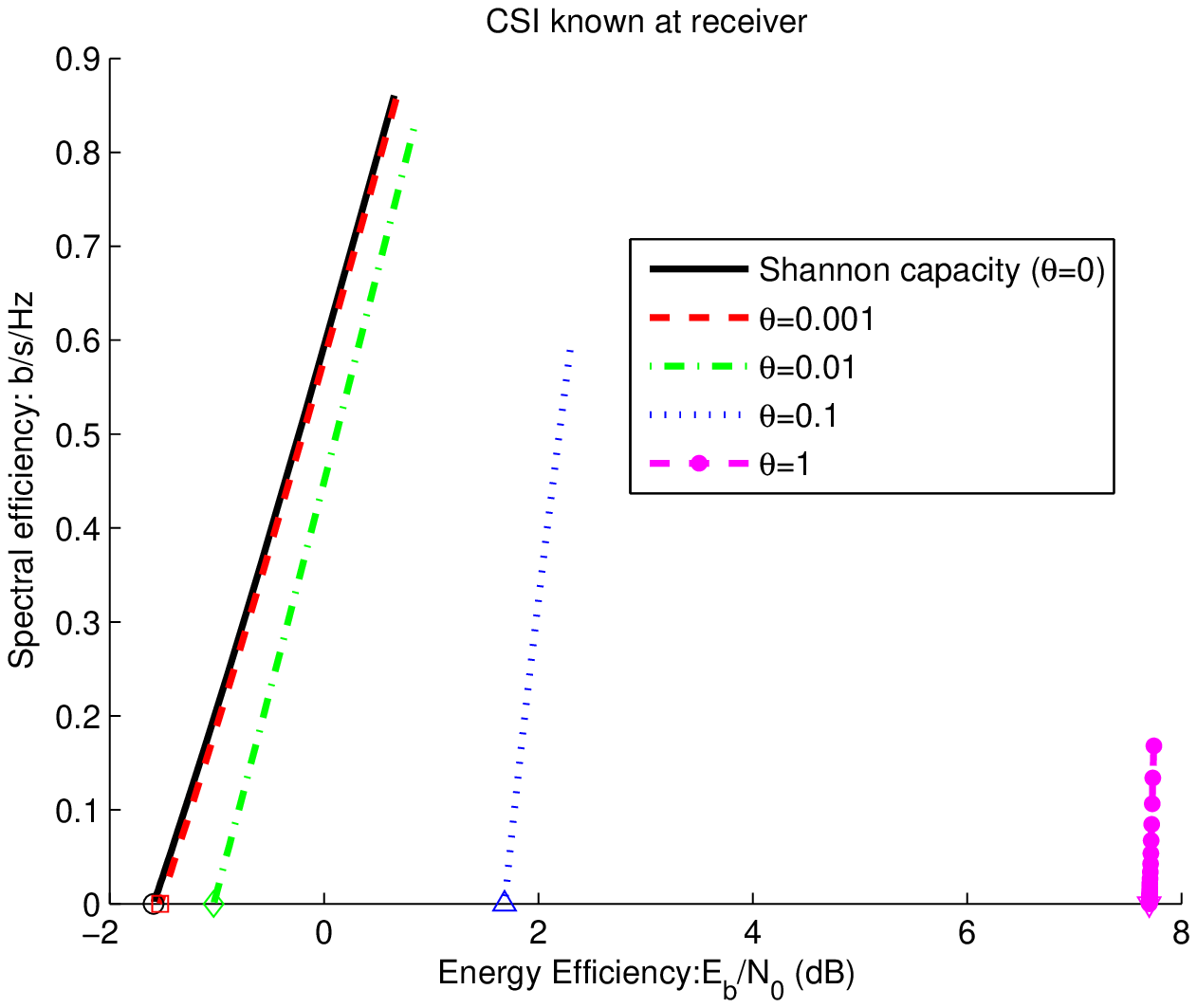}
\caption{Spectral efficiency vs. $E_{b}/N_{0}$ in the Rayleigh
fading channel with fixed $\Pb$; CSI known at the receiver
only.}\label{fig:5}
\end{center}
\end{figure}

It is interesting to note that unlike the low-power regime results,
we now have
\begin{gather}
\frac{E_b}{N_0}_{\tmin}=\frac{-\frac{\theta
T\Pb}{N_0}}{\log_e\E\{e^{-\frac{\theta T\Pb}{N_{0}\log_e2}z}\}} \ge
\frac{-\frac{\theta T\Pb}{N_0}}{\E\{\log_e e^{-\frac{\theta
T\Pb}{N_{0}\log_e2}z}\}} = \frac{\log_e 2}{\E\{z\}} \nonumber
\end{gather}
where Jensen's inequality is used. Therefore, we will be operating
above $-1.59$ dB unless there are no QoS constraints and hence
$\theta = 0$. For the Rayleigh channel, we can specialize
(\ref{eq:ebminwbr}) and (\ref{eq:slopewbr}) to obtain
\begin{equation}\label{ebminbrx}
\frac{E_b}{N_0}_{\textrm{min}}=\frac{\frac{\theta
T\Pb}{N_0}}{\log_e(1+\frac{\theta T \Pb}{N_0\log_e2})} \quad
\text{and} \quad \mathcal{S}_0=\bigg(\frac{N_0\log_e2}{\theta T
\Pb}\log_e(1+\frac{\theta T \Pb}{N_0\log_e2})+\log_e(1+\frac{\theta
T \Pb}{N_0\log_e{2}})\bigg)^2.
\end{equation}
It can be easily seen that in the Rayleigh channel, the minimum bit
energy monotonically increases with increasing $\theta$. Fig.
\ref{fig:5} plots the spectral efficiency curves as a function of
bit energy in the Rayleigh channel. In all the curves, we set
$\Pb/N_{0}=10^4$. We immediately observe that more stringent QoS
constraints and hence higher values of $\theta$ lead to higher
minimum bit energy values and also higher energy requirements at
other nonzero spectral efficiencies. The wideband slope values are
found to be equal to
$\mathcal{S}_0=\{1.0288,1.2817,3.3401,12.3484\}$ for
$\theta=\{0.001,0.01,0.1,1\}$, respectively.

\begin{figure}
\begin{center}
\includegraphics[width=0.65\textwidth]{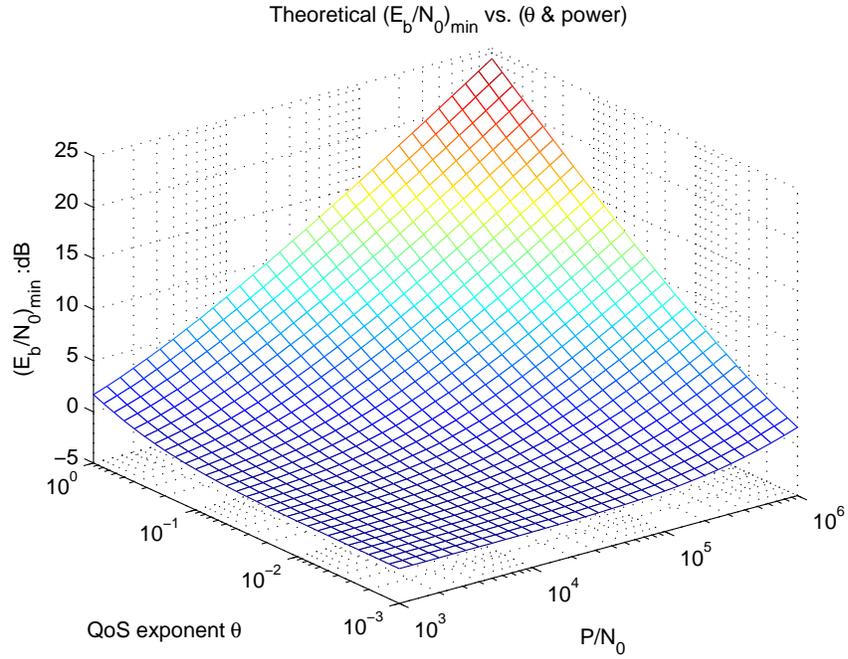}
\caption{$\frac{E_b}{N_0}_{\textrm{min}}$ vs. $\theta$ and $\Pb/N_0$
in the Rayleigh fading channel; CSI known at the receiver
only.}\label{fig:10}
\end{center}
\end{figure}

We finally note that $\frac{E_b}{N_0}_{\tmin}$ and $\mathcal{S}_0$
now depend on $\theta$ and $\frac{\Pb}{N_0}$. Fig. \ref{fig:10}
plots $\frac{E_b}{N_0}_\tmin$ as a function of these two parameters.
Probing into the inherent relationships among these parameters can
give us some interesting results, which are helpful in designing
wireless networks. For instance, for some $\Pb/N_0$ required to
achieve some specific transmission rate, we can find the most
stringent QoS guarantee possible while attaining a certain
efficiency in the usage of energy, or if a QoS requirement $\theta$
is specified, we can find the minimum power $\Pb$ to achieve a
specific bit energy.


\subsection{CSI at both the Transmitter and Receiver}
To analyze $\frac{E_b}{N_0}_{\tmin}$ in this case, we initially
obtain the following result and identify the limiting value of the
threshold $\alpha$ as the bandwidth increases to infinity.
\begin{Lem}
In wideband regime, the threshold $\alpha$ in the optimal power
adaptation scheme (\ref{power}) satisfies\vspace{-.3cm}
\begin{equation}\label{alphast}
\lim_{\zeta \rightarrow0}\alpha(\zeta)=\alpha^{\ast}
\end{equation}
where $\alpha^{\ast}$ is the solution to \vspace{-.3cm}
\begin{equation}\label{alpha}
\E\left\{\left[\log_e\left(\frac{z}{\alpha^{\ast}}\right)\frac{1}{z}\right]\tau(\alpha^{\ast})\right\}=\frac{\theta
T\Pb}{N_{0}\log_e{2}}.
\end{equation}
Moreover, for $\theta > 0$, $\sta < \infty$.
\end{Lem}
\emph{Proof}: Recall from (\ref{pcon}) that the optimal power
adaptation rule should satisfy the average power constraint:
\begin{align}\label{bcon}
\tsnr = \frac{\Pb \zeta}{N_0}
&=\E\bigg\{\Big(\frac{1}{\alpha^{\frac{1}{\beta+1}}z^{\frac{\beta}{\beta+1}}}-\frac{1}{z}\Big)\tau(\alpha)\bigg\}
=\E\bigg\{\bigg[\bigg(\Big(\frac{z}{\alpha}\Big)^{\frac{1}{\beta+1}}-1\bigg)\frac{1}{z}\bigg]\tau(\alpha)\bigg\}
\end{align}
where $\beta=\frac{\theta TB}{\log_e2}=\frac{\theta
T}{\zeta\log_e2}$. For the case in which $\theta = 0$, if we let
$\zeta \to 0$, we obtain from (\ref{bcon}) that
\begin{gather}
0 = \E \left\{ \left[ \left(\frac{z}{\sta}-1
\right)\frac{1}{z}\right]\tau(\sta)\right\}
\end{gather}
where $\sta = \lim_{\zeta \to 0} \alpha(\zeta)$. Using the fact that
$\log_e x \le x - 1$ for $x \ge 1$, we have $\log_e \left(
\frac{z}{\sta}\right)  \le \frac{z}{\sta} - 1$ for $z \ge \sta$
which implies that
\begin{gather}
0 \le
\E\left\{\left[\log_e\left(\frac{z}{\alpha^{\ast}}\right)\frac{1}{z}\right]\tau(\alpha^{\ast})\right\}
\le \E \left\{ \left[ \left(\frac{z}{\sta}-1
\right)\frac{1}{z}\right]\tau(\sta)\right\} = 0 \implies
\E\left\{\left[\log_e\left(\frac{z}{\alpha^{\ast}}\right)\frac{1}{z}\right]\tau(\alpha^{\ast})\right\}
= 0 \nonumber
\end{gather}
proving (\ref{alpha}) for the case of $\theta = 0$.

In the following, we assume $\theta > 0$. We first define
$g(\zeta)=\left(\frac{z}{\alpha}\right)^{\frac{1}{\beta + 1}} =
\left(\frac{z}{\alpha}\right)^{\frac{\zeta\log_e2}{\zeta\log_e2+\theta
T}}$ and take the logarithm of both sides to obtain
\begin{align}
\log_e g(\zeta)=\frac{\zeta\log_e2}{\zeta\log_e2+\theta
T}\log_e\frac{z}{\alpha}.
\end{align}
Differentiation over both sides leads to
\begin{align}\label{eq:dotg}
\frac{\dot{g}(\zeta)}{g(\zeta)}=\frac{\theta
T\log_e2}{(\zeta\log_e2+\theta
T)^2}\log_e\frac{z}{\alpha}-\frac{\zeta\log_e2}{\zeta\log_e2+\theta
T}\frac{\dot{\alpha}}{\alpha}
\end{align}
where $\dot{g}$ and $\dot{\alpha}$ denote the first derivatives $g$
and $\alpha$, respectively, with respect to $\zeta$. 
Noting that $g(0)=1$, we can see from (\ref{eq:dotg}) that as $\zeta
\to 0$, we have
\begin{equation}\label{doty}
\dot{g}(0)=\frac{\log_e2}{\theta T}\log_e\frac{z}{\sta}
\end{equation}
where $\sta=\lim_{\zeta\rightarrow0}\alpha(\zeta)$. For small values
of $\zeta$, the function $g$ admits the following Taylor series:
\begin{equation}\label{temptayl}
g(\zeta) = \left(\frac{z}{\alpha}\right)^{\frac{1}{\beta+1}}
=g(0)+\dot{g}(0)\zeta+o(\zeta) = 1+\dot{g}(0)\zeta+o(\zeta).
\end{equation}
Therefore, we have
\begin{equation}
\Big(\frac{z}{\alpha}\Big)^{\frac{1}{\beta+1}}-1=\frac{\log_e2}{\theta
T}\log_e\left(\frac{z}{\sta}\right)\zeta+o(\zeta).
\end{equation}
Then, from (\ref{bcon}), we can write
\begin{equation} \label{eq:avgpowerconst}
\tsnr=\E\left\{\left[\left(\frac{\log_e2}{\theta
T}\log_e\left(\frac{z}{\alpha}\right)\zeta+o(\zeta)\right)\,\frac{1}{z}\right]\tau(\alpha)\right\}.
\end{equation}
If we divide both sides of (\ref{eq:avgpowerconst}) by $\tsnr =
\frac{\Pb \zeta}{N_0}$ and let $\zeta \to 0$, we obtain
\begin{equation}\label{constaintp}
\lim_{\zeta\rightarrow0}\frac{\tsnr}{\tsnr}=
\lim_{\zeta\rightarrow0}\frac{\tsnr}{\frac{\Pb \zeta}{N_0}}= 1 =
\frac{N_{0}\log_e2}{\theta
T\Pb}\E\left\{\left[\log_e\left(\frac{z}{\sta}\right)\frac{1}{z}\right]\tau(\sta)\right\}
\end{equation}
from which we conclude that
$\E\left\{\left[\log_e\left(\frac{z}{\alpha^{\ast}}\right)\frac{1}{z}\right]\tau(\alpha^{\ast})\right\}=\frac{\theta
T\Pb}{N_{0}\log_e{2}}$, proving (\ref{alpha}) for $\theta > 0$. 

Using the fact that
$\log_e\left(\frac{z}{\alpha}\right)<\frac{z}{\alpha}$ for $z \ge
0$, we can write
\begin{equation}\label{eq:finitethresh}
0\leq
\E\left\{\left[\log_e\left(\frac{z}{\alpha}\right)\frac{1}{z}\right]\tau(\alpha)\right\}
\leq \E\bigg\{\frac{1}{\alpha}\tau(\alpha)\bigg\} \le
\frac{1}{\alpha}.
\end{equation}
Assume now that $\lim_{\zeta \to 0} \alpha(\zeta) = \sta = \infty$.
Then, the rightmost side of (\ref{eq:finitethresh}) becomes zero in
the limit as $\zeta \to 0$ which implies that
$\E\left\{\left[\log_e\left(\frac{z}{\sta}\right)\frac{1}{z}\right]\tau(\sta)\right\}
= 0$. From (\ref{alpha}), this is clearly not possible for $\theta
> 0$. Hence, we have proved that $\sta < \infty$ when $\theta > 0$.
\hfill$\square$

\begin{figure}
\begin{center}
\includegraphics[width=0.65\textwidth]{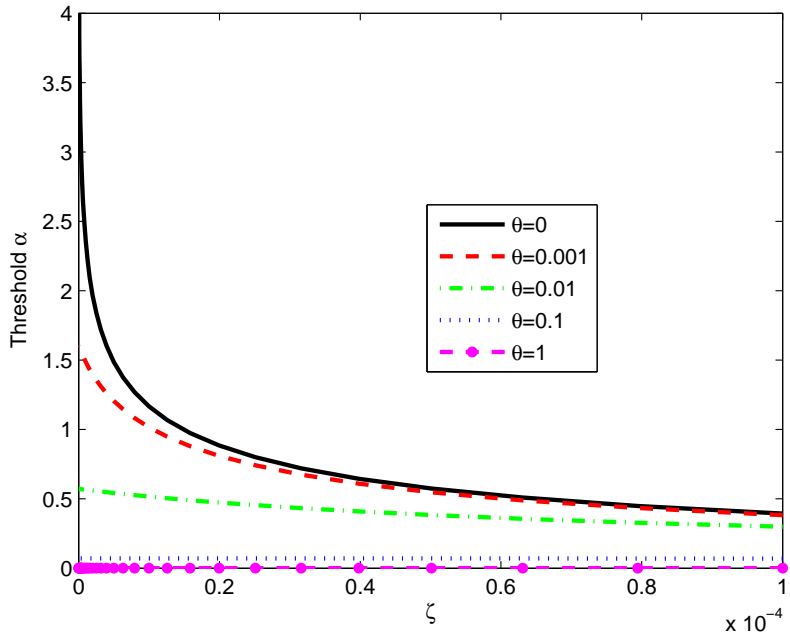}
\caption{Threshold of channel gain $\alpha$ vs. $\zeta$ in the
Rayleigh fading channel; CSI known at the transmitter and
receiver.}\label{fig:7}
\end{center}
\end{figure}

\emph{Remark:} As noted before, wideband and low-power regimes are
equivalent when $\theta  = 0$. Hence, as in the proof of Theorem
\ref{theo:lowpowercsirt}, we can easily see in the wideband regime
that the threshold $\alpha$ approaches the maximum fading value
$z_{\max}$ as $\zeta \to 0$ when $\theta = 0$. Hence, for fading
distributions with unbounded support, $\alpha \to \infty$ with
vanishing $\zeta$. The threshold being very large means that the
transmitter waits sufficiently long until the fading assumes very
large values and becomes favorable. That is how arbitrarily small
bit energy values can be attained. However, in the presence of QoS
constraints, arbitrarily long waiting times will not be permitted.
As a result, $\alpha$ approaches a finite value (i.e., $\sta <
\infty$) as $\zeta \to 0$ when $\theta
> 0$. Moreover, from (\ref{alpha}), we can immediately note that as
$\theta$ increases, $\sta$ has to decrease. This fact can also be
observed in Fig. \ref{fig:7} in which $\alpha$ vs. $\zeta$ is
plotted in the Rayleigh fading channel. Consequently, arbitrarily
small bit energy values will no longer be possible when $\theta > 0$
as will be shown in Theorem \ref{theo:widebandcsirt}.

The spectral efficiency with optimal power adaptation is now given
by
\begin{equation}\label{ectxrxb}
\C_E(\zeta)=-\frac{\zeta}{\theta T}\log_e\Big(F(\alpha)
+\E\Big\{\Big(\frac{z}{\alpha}\Big)^{-\frac{\theta T}{\theta
T+\zeta\log_e{2}}}\tau(\alpha)\Big\}\Big)
\end{equation}
where again $F(\alpha)=\E\{1\{z<\alpha\}\}$ and $\tau(\alpha) =
1\{\tau \ge \alpha\}$. 
\begin{Lem}\label{theo:widebandcsirt}
When both the receiver and transmitter have CSI, the minimum bit
energy and wideband slope in the wideband regime are given by
\begin{align}\label{resulttxrxb}
\frac{E_b}{N_0}_{\min}=-\frac{\theta T\Pb}{N_{0}\log_e\xi} \text{
and } \mathcal{S}_0=\frac{\xi(\log_e\xi)^{2} \log_e2}{\theta
T(\frac{\Pb\sta}{N_0}+\dot{\alpha}(0)\E\{\frac{1}{z}\tau(\sta)\})}
\end{align}
where
$\xi=F(\alpha^{\ast})+\E\{\frac{\alpha^{\ast}}{z}\tau(\alpha^{\ast})\}$,
and $\dot{\alpha}(0)$ is the derivative of $\alpha$ with respect to
$\zeta$, evaluated at $\zeta = 0$. 
\end{Lem}

\emph{Proof}: Substituting (\ref{ectxrxb}) into (\ref{ebminb}) leads
to
\begin{align}\label{ebminbtx}
\frac{E_b}{N_0}_{\min}&=\lim_{\zeta\rightarrow0}\frac{\Pb\zeta/N_0}{-\frac{\zeta}{\theta
T}\log_e\Big(F(\alpha)+\E\Big\{\Big(\frac{z}{\alpha}\Big)^{-\frac{\theta
T}{\theta T+\zeta\log_e{2}}}\tau(\alpha)\Big\}\Big)}=-\frac{\theta
T\Pb}{N_{0}\log_e\Big(F(\alpha^{\ast})+\E\Big\{\frac{\alpha^{\ast}}{z}\tau(\alpha^{\ast})\Big\}\Big)}.
\end{align}
After denoting
$\xi=F(\alpha^{\ast})+\E\{\frac{\alpha^{\ast}}{z}\tau(\alpha^{\ast})\}$,
we obtain the expression for minimum bit energy in
(\ref{resulttxrxb}).

Meanwhile, $\C_E(\zeta)$ has the following Taylor series expansion
up to second order:
\begin{equation}
\C_E(\zeta)=\dot{\C}_E(0)\zeta+\frac{1}{2}\ddot{\C}_E(0)\zeta^{2}+o(\zeta^2).
\end{equation}
Therefore, the second derivative of $\C_E$ with respect to $\zeta$
at $\zeta = 0$ can be computed from
\begin{equation}
\ddot{\C}_E(0)=2\lim_{\zeta\rightarrow0}\frac{\C_E(\zeta)-\dot{\C_E}(0)\zeta}{\zeta^2}.
\end{equation}
From the derivation of (\ref{ebminbtx}) and (\ref{ebminb}), we know
that
\begin{align}\label{dotb}
 \dot{\C}_E(0)=-\frac{1}{\theta
T}\log_e\Big(F(\alpha^{\ast})+\E\Big\{\frac{\alpha^{\ast}}{z}\tau(\alpha^{\ast})\Big\}\Big).
\end{align}
Then, \vspace{-0.55cm}
\begin{align}
\ddot{\C}_E(0)&=2\lim_{\zeta\rightarrow0}\frac{-\frac{\zeta}{\theta
T}\log_e\Big(F(\alpha)
+\E\Big\{\Big(\frac{z}{\alpha}\Big)^{-\frac{\theta T}{\theta
T+\zeta\log_e{2}}}\tau(\alpha)\Big\}\Big)+\frac{\zeta}{\theta
T}\log_e\Big(F(\alpha^{\ast})+\E\Big\{\frac{\sta}{z}\tau(\sta)\Big\}\Big)}{\zeta^2}\\
&=-\frac{2}{\theta
T}\lim_{\zeta\rightarrow0}\frac{\log_e\frac{F(\alpha)+\E\Big\{\big(\frac{\alpha}{z}\big)^{\frac{\theta
T}{\theta
T+\zeta\log_e2}}\tau(\alpha)\Big\}}{F(\sta)+\E\Big\{\frac{\sta}{z}\tau(\sta)\Big\}}}{\zeta}\\
&=-\frac{2}{\theta
T}\lim_{\zeta\rightarrow0}\frac{\E\bigg\{(\frac{\alpha}{z})^{\frac{\theta
T}{\theta T+\zeta \log_e2}} \Big(-\frac{\theta T\log_e2}{(\theta
T+\zeta\log_e2)^2}\log_e\big(\frac{\alpha}{z}\big)+\frac{\theta
T\dot{\alpha}}{(\theta T+\zeta\log_e2)\alpha}\Big)\tau(\alpha)\bigg\}}{F(\sta)+\E\Big\{\frac{\sta}{z}\tau(\sta)\Big\}} \label{eq:secondderiv}\\
&=-\frac{2\log_e2}{(\theta
T)^2}\frac{\E\bigg\{\frac{\sta}{z}\log_e\Big(\frac{z}{\sta}\Big)\tau(\sta)\bigg\}+\frac{\theta
T\dot{\alpha}(0)}{\log_e2}\E\{\frac{1}{z}\tau(\sta)\}}{F(\sta)+\E\Big\{\frac{\sta}{z}\tau(\sta)\Big\}},\label{ddotb}
\end{align}
where $\dot{\alpha}$ is the derivative of $\alpha$ with respect to
$\zeta$. Above, (\ref{eq:secondderiv}) is obtained by using
L'Hospital's Rule.  Evaluating (\ref{ecslope}) with (\ref{dotb}) and
(\ref{ddotb}), and combining with the result in (\ref{alpha}), we
obtain the expression for $\mathcal {S}_0$ in
(\ref{resulttxrxb}).\hfill $\square$

\begin{figure}
\begin{center}
\includegraphics[width=0.65\textwidth]{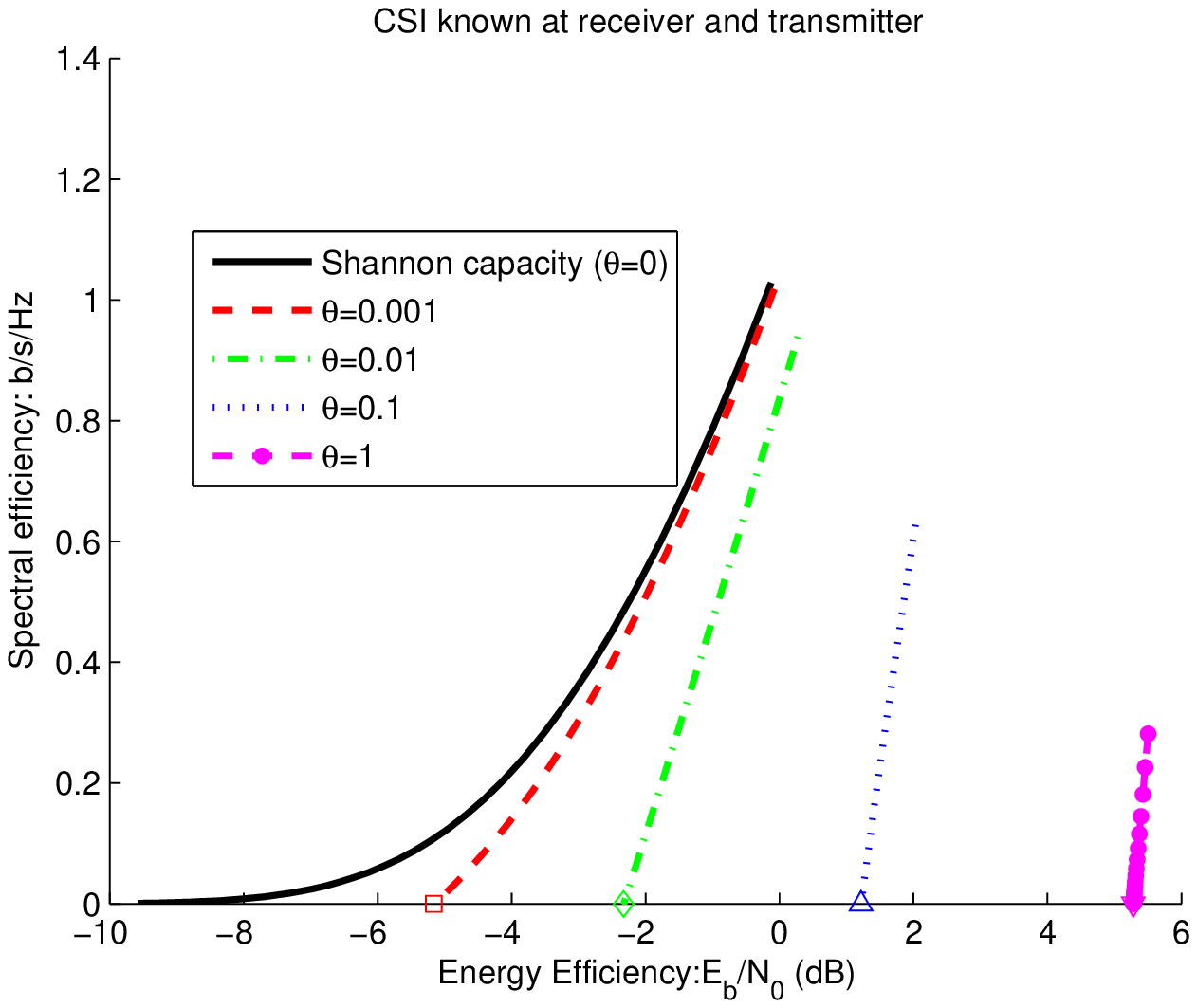}
\caption{Spectral efficiency vs. $E_{b}/N_{0}$ in the Rayleigh
fading channel with fixed $ \frac{\Pb}{N_0} = 10^4$; CSI known at
the transmitter and receiver.}\label{fig:8}
\end{center}
\end{figure}

\begin{figure}
\begin{center}
\includegraphics[width=0.65\textwidth]{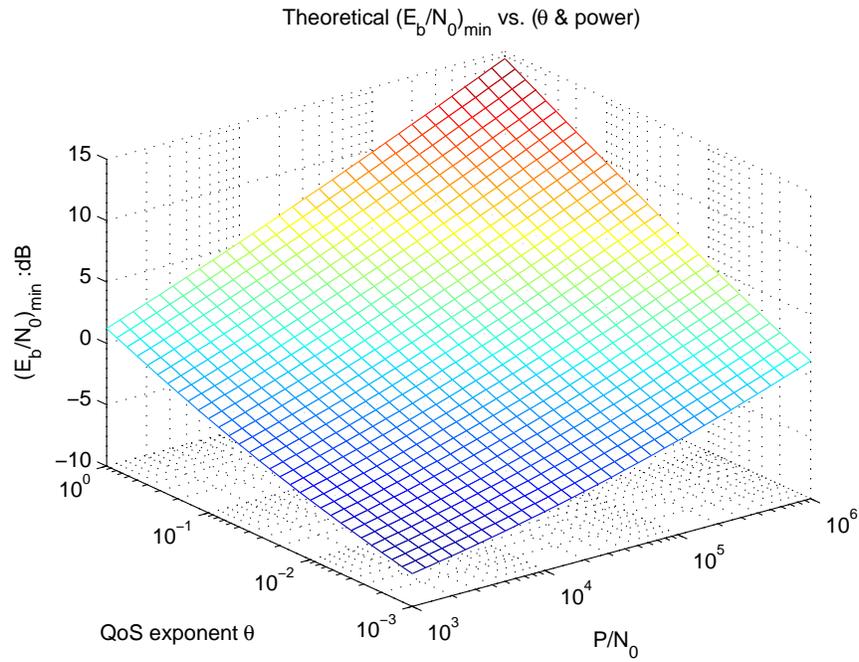}
\caption{$\frac{E_b}{N_0}_{\min}$ vs. $\theta$ and $\Pb/N_0$ in the
Rayleigh fading channel; CSI known at the transmitter and
receiver.}\label{fig:11}
\end{center}
\end{figure}

\begin{figure}
\begin{center}
\includegraphics[width=0.65\textwidth]{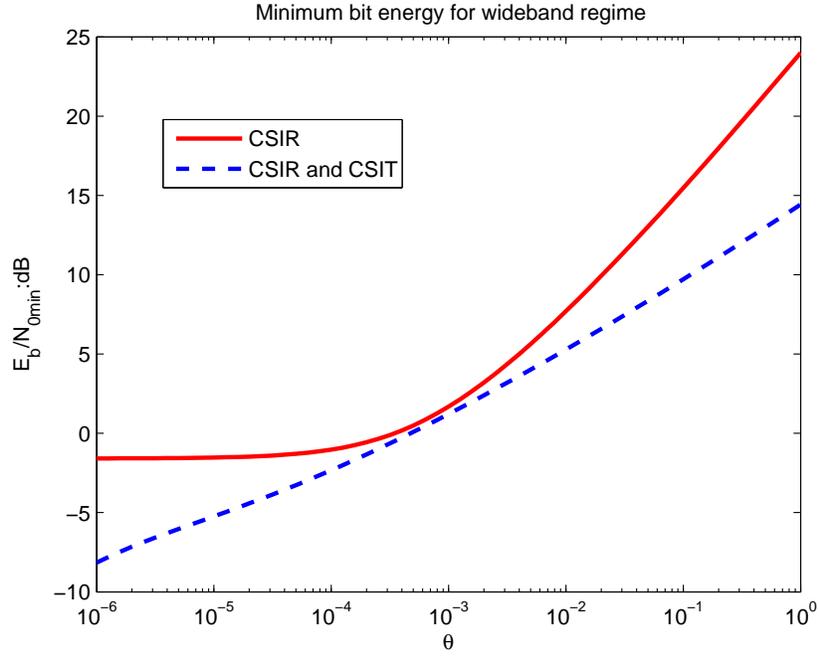}
\caption{$\frac{E_b}{N_0}_{\min}$ vs. $\theta$ in the Rayleigh
fading channel. $\frac{\Pb}{N_0} = 10^4$.}\label{fig:ebr_rev}
\end{center}
\end{figure}

It is interesting to note that the minimum bit energy is strictly
greater than zero for $\theta > 0$. Hence, we see a stark difference
between the wideband regime and low-power regime in which the
minimum bit energy is zero for fading distributions with unbounded
support. 
Fig. \ref{fig:8} plots the spectral efficiency curves in the
Rayleigh fading channel and is in perfect agreement with the
theoretical results. Obviously, the plots are drastically different
from those in the low-power regime (Fig. \ref{fig:4}) where all
curves approach $-\infty$ as the spectral efficiency decreases. In
Fig. \ref{fig:8}, the minimum bit energy is finite for the cases in
which $\theta > 0$. The wideband slope values are computed to be
equal to $\mathcal{S}_0=\{0.3081,1.0455, 2.5758,4.1869\}$. Fig.
\ref{fig:11} plots the  $\frac{E_b}{N_0}_{\min}$ as a function of
$\theta$
and $\Pb/N_0$. 
Generally speaking, due to power and rate adaptation,
$\frac{E_b}{N_0}_{\min}$ in this case is smaller compared to that in
the case in which only the receiver has CSI. This can be observed in
Fig. \ref{fig:ebr_rev} where the minimum bit energies are compared.
From Fig. \ref{fig:ebr_rev}, we note that the presence of CSI at the
transmitter is especially beneficial for very small and also large
values of $\theta$. While the bit energy in the CSIR case approaches
$-1.59$ dB with vanishing $\theta$, it decreases to $-\infty$ dB
when also the transmitter knows the channel. On the other hand, when
$\theta \approx 10^{-3}$, we interestingly observe that there is not
much to be gained in terms of the minimum bit energy by having CSI
at the transmitter. For $\theta > 10^{-3}$, we again start having
improvements with the presence of CSIT.

\begin{figure}
\begin{center}
\includegraphics[width=0.65\textwidth]{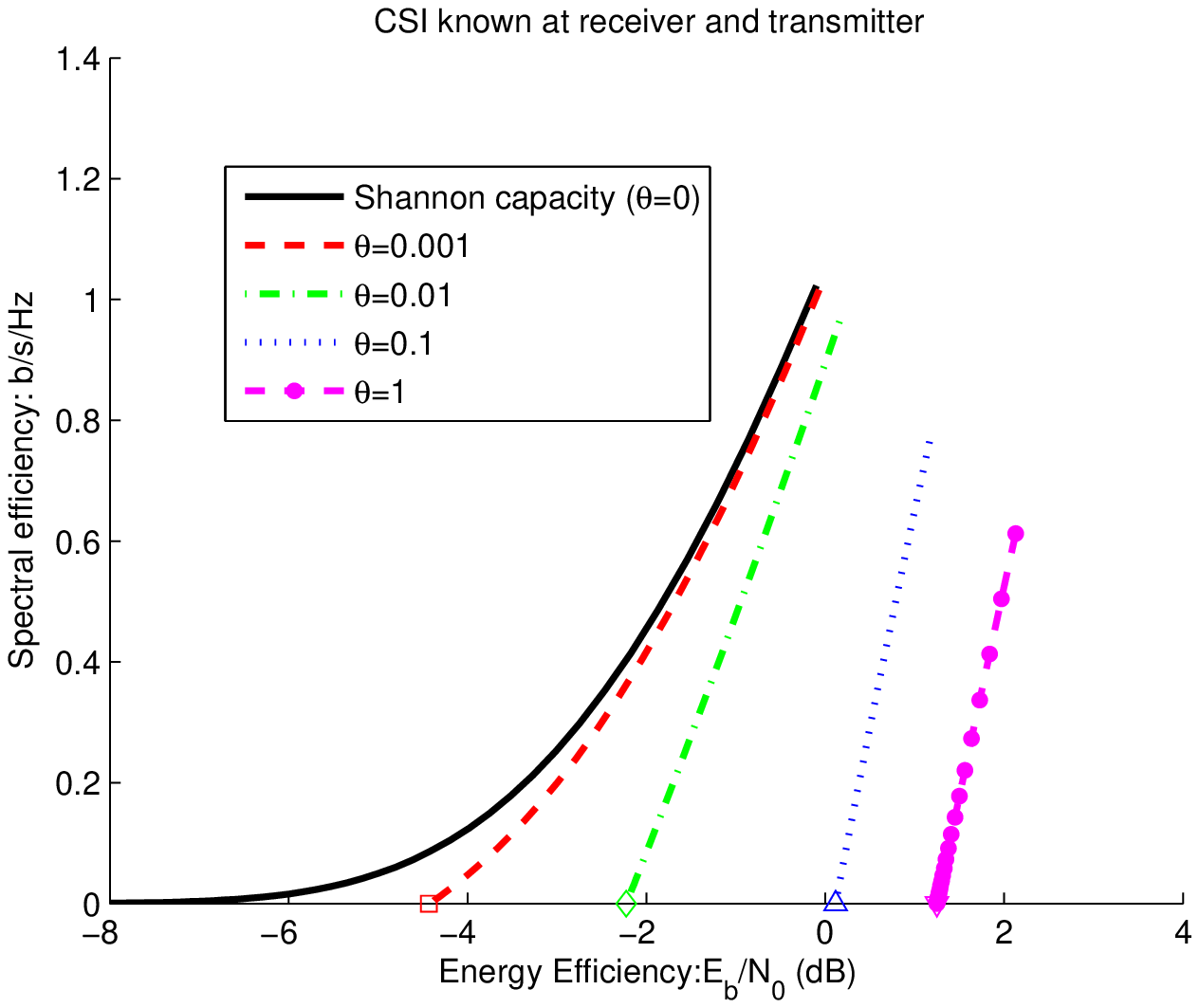}
\caption{Spectral efficiency vs. $E_{b}/N_{0}$ in the Nakagami-$m$
fading channel with $m = 2$; $\frac{\Pb}{N_0} = 10^4$ ;CSI known at
the transmitter and receiver.}\label{fig:txnaka_2}
\end{center}
\end{figure}

Throughout the paper, numerical results are provided for the
Rayleigh fading channel. However, note that the theoretical results
hold for general stationary and ergodic fading processes. Hence,
other fading distributions can easily be accommodated as well. In
Fig. \ref{fig:txnaka_2}, we plot the spectral efficiency vs. bit
energy curves for the Nakagami-$m$ fading channel with $m = 2$.

\section{Conclusion}

In this paper, we have analyzed the energy efficiency in fading
channel under QoS constraints by considering the effective capacity
as a measure of the maximum throughput under certain statistical QoS
constraints, and analyzing the bit energy levels. Our analysis has
provided a characterization of the energy-bandwidth-delay tradeoff.
In particular, we have investigated the spectral efficiency vs. bit
energy tradeoff in the low-power and wideband regimes under QoS
constraints. We have elaborated the analysis under two scenarios:
perfect CSI available at the receiver and perfect CSI available at
both the receiver and transmitter. We have obtained expressions for
the minimum bit energy and wideband slope. Through this analysis, we
have quantified the increased energy requirements in the presence of
delay-QoS constraints. While the bit energy levels in the low-power
regime can approach those that can be attained in the absence of QoS
constraints, we have shown that strictly higher bit energy values
are needed in the wideband regime. For instance, we have shown that
when both the transmitter and receiver has perfect CSI,
$\frac{E_b}{N_0}_{\min} > 0$ in the wideband regime for $\theta
> 0$ while $\frac{E_b}{N_0}_{\min} = 0$ if $\theta =
0$ for fading distributions with unbounded support. 
We have also provided numerical results by considering the Rayleigh
and Nakagami fading channels and verified the theoretical
conclusions.

\appendices

\section{} \label{app:2}

Considering (\ref{rxecb}), we denote
\begin{equation}
C_E(\zeta)=\frac{\C_E(\zeta)}{\zeta}=-\frac{1}{\theta
T}\log_e{\mathbb{E}\{e^{-\frac{\theta
T}{\zeta}\log_2(1+\frac{\Pb\zeta}{N_0}z)}\}}.
\end{equation}
The first derivative of $C_E(\zeta)$ with respect to $\zeta$ is
given by
\begin{equation}
\dot{C}_E(\zeta)=-\frac{1}{\zeta^2\log_e2}\frac{\E\{e^{-\frac{\theta
T}{\zeta}\log_e{2}(1+\frac{\Pb\zeta}{N_0}z)}\big[\log_{e}(1+\frac{\Pb\zeta}{N_0}z)-\frac{\frac{\Pb\zeta}{N_0}z}{1+\frac{\Pb\zeta}{N_0}z}\big]\}}{\E\{e^{-\frac{\theta
T}{\zeta}\log_{2}(1+\frac{\Pb\zeta}{N_0}z)}\}}.
\end{equation}
We let $\nu=\frac{\Pb\zeta}{N_0}z\geq0$, and define
$y(\nu)=\log_e(1+\nu)-\frac{\nu}{1+\nu}$, where $y(0)=0$. It can be
easily seen that $\dot{y}=\frac{\nu}{(1+\nu)^2}\geq0$, so
$y(\nu)\geq0$ holds for all $\nu$. Then, we immediately observe that
$\dot{C}_E(\zeta)\leq0$. Therefore, $\frac{\C_E(\zeta)}{\zeta}$
monotonically increases with \emph{decreasing} $\zeta$.
\end{spacing}
\begin{spacing}{1.4}

\end{spacing}

\newpage

%
%


\begin{thebibliography}{99}
\vspace{-.04cm}

\bibitem{Biglieri} E. Biglieri, J. Proakis, and S. Shamai (Shitz), ``Fading channels: Information-theoretic
and communications aspects," \emph{IEEE Trans. Inform. Theory},
vol.~44, pp.~2619-2692, Oct. 1998.

\bibitem{sergio} S. Verd$\acute{\text{u}}$, ``Spectral efficiency in the wideband regime,''
 \emph{IEEE Trans. Inform. Theory}, vol.48, no.6~pp.1319-1343.
 Jun.2002.

\bibitem{Ephremides} A. Ephremides and B. Hajek, ``Information
theory and communication networks: An unconsummated union,"
\emph{IEEE Trans. Inform. Theory}, vol.~44, pp.~2416-2434, Oct.
1998.

\bibitem{Ozarow} L. Ozarow, S. Shamai (Shitz), and A. Wyner, ``Information theoretic considerations
for cellular mobile radio,'' \emph{IEEE Trans. Veh. Technol.},
vol.~43, pp.~359-378, May 1994.

\bibitem{delay} S. V. Hanly and D.N.C Tse, ``Multiaccess fading
channels-part II: delay-limited capacities,'' \emph{IEEE Trans.
Inform. Theory}, vol.44, no.7, pp.2816-2831. Nov. 1998.

\bibitem{Telatar} I. E. Telatar and R. G. Gallager, ``Combining queueing theory with information theory for multiaccess,"
\emph{IEEE J. Select. Areas Commun.}, vol.~13, pp.~963-969, Aug.
1995.

\bibitem{Berry} R. A. Berry and R. G. Gallager, ``Communication over fading channels with delay constraints,"
\emph{IEEE Trans. Inform. Theory}, vol.~48, pp.~1135-1149, May 2002.

\bibitem{Neely} M. J. Neely, ``Optimal energy and delay tradeoffs for multiuser wireless downlinks,"
\emph{IEEE Trans. Inform. Theory}, vol.~53, pp.~3095-3113, Sept.
2007.

\bibitem{Kelly} F. Kelly, ``Notes on effective bandwidths," Stochastic Networks: Theory and
Applications, Royal Statistical Society Lecture Notes Series, 4.,
pp:~141-168, Oxford University Press, 1996.

\bibitem{chang} C.-S. Chang, ``Stability, queue length, and delay of deterministic and stochastic queuing
networks,'' \emph{IEEE Trans. Auto. Control}, vol. 39, no. 5, pp.
913-931, May 1994.

\bibitem{changEB} C.-S. Chang, ``Effective bandwidth in high-speed
digital networks," \emph{IEEE J. Sel. Areas Commun.}, vol.13, pp.
1091-1100, Aug. 1995.

\bibitem{ChangZajic} C.-S. Chang and T. Zajic, ``Effective bandwidths
of departure processes from queues  with time varying capacities,"
INFOCOM, 1995.

\bibitem{dapeng} D. Wu and R. Negi ``Effective capacity: a wireless link model for support of quality of service,''
 \emph{IEEE Trans. Wireless Commun.}, vol.2,no. 4, pp.630-643. July
 2003.

\bibitem{dapengw} D. Wu and R. Negi, ``Effective capacity-based quality of service measures for wireless networks,''
 \emph{Proc. First
International Conference on Broadband Networks, 2004}, pp. 527-536.

\bibitem{dw} D. Wu and R. Negi, ``Downlink scheduling in a cellular network for quality-of-service
assurance,'' \emph{IEEE Trans. Veh. Technol.}, vol.53, no.5, pp.
1547-1557, Sep., 2004.

\bibitem{uti} D. Wu and R. Negi, ``Utilizing multiuser diversity for efficient support of quality of service over a fading
channel,'' \emph{IEEE Trans. Veh. Technol.}, vol.49, pp. 1073-1096,
May 2003.

\bibitem{jia} J. Tang and X. Zhang, ``Quality-of-Service Driven Power and Rate Adaptation over Wireless
Links,'' \emph{IEEE Trans. Wireless Commun.}, vol. 6, no. 8,
pp.3058-3068, Aug. 2007.

\bibitem{multi} J. Tang and X. Zhang, ``Quality-of-Service Driven Power and Rate Adaptation for Multichannel Communications over Wireless
Links,'' \emph{IEEE Trans. Wireless Commun.}, vol. 6, no. 12,
pp.4349-4360, Dec. 2007.

\bibitem{tangzhangcross} J. Tang and X. Zhang, ``Cross-layer modeling for quality of service guarantees over
wireless links,'' \emph{IEEE Trans. Wireless Commun.}, vol. 6, no.
12, pp.4504-4512, Dec. 2007.

\bibitem{tangzhangcross2} J. Tang and X. Zhang, ``Cross-layer-model based adaptive resource allocation
for statistical QoS guarantees in mobile wireless networks,''
\emph{IEEE Trans. Wireless Commun.}, vol. 7, pp.2318-2328, June
2008.



\bibitem{finite} L. Liu, P. Parag, J. Tang, W.-Y. Chen and J.-F.
Chamberland, ``Resource allocation and quality of service evaluation
for wireless communication systems using fluid models,'' \emph{IEEE
Trans. Inform. Theory}, vol. 53, no. 5, pp. 1767-1777, May 2007.


%

\bibitem{cdma} S. Shamai, S. Verd$\acute{\text{u}}$, ``The impact of
frequency-flat fading on the spectral efficiency of CDMA,''
\emph{IEEE Trans. Inform. Theory}, vol. 47, no. 4, pp. 1302-1327,
May 2001.

\bibitem{sha} S. Borade and L. Zheng, ``Wideband fading channels with feedback,''
42nd Allerton Annual Conference on Communication, Control and
Computing, October 2004.

\bibitem{Luby} M. Luby, ``LT codes," Proc. 43rd Ann. IEEE Symp. Found. Comp.
Sci., 2002, pp. 271–280.

\bibitem{Shok} A. Shokrollahi, ``Raptor codes," \emph{IEEE Trans. Inform. Theory}, vol.~52, pp.~2551-2567,
June 2006.

\bibitem{Castura1} J. Casture and Y. Mao, ``Rateless coding and relay
networks," \emph{IEEE Signal Process. Mag.}, vol.~24, pp. 27-35,
Sept. 2007.

\bibitem{Castura2} J. Casture and Y. Mao, ``Rateless coding over fading channels,"
 \emph{IEEE Comm. Letters}, vol.~10, pp. 46-48,
Jan. 2006.


\bibitem{Cover} T. M. Cover and J. A. Thomas, \emph{Elements of Information
Theory.} New York: Wiley, 1991.

\bibitem{Changbook} C.-S. Chang, \emph{Performance Guarantees in Communication
Networks}, Springer-Verlag, 2000.

\bibitem{convex} S. Boyd and L. Vandenberghe, \emph{Convex
Optimization}, Cambridge University Press, 2004.

\bibitem{Garg} V. K. Garg, \emph{Wireless Communications and
Networking}, Elsevier, 2007.

\bibitem{rapp} T. S. Rappaport, \emph{Wireless Communications: Principles and Practice},
 2nd ed. Prentice Hall PTR, 2001.

\end{thebibliography}
\end{document}